\documentclass[abstract]{aa}
\usepackage[T1]{fontenc}
\usepackage{ae,aecompl}
\usepackage{graphicx}
\usepackage{txfonts}
\usepackage{amsmath}
\usepackage{amssymb}
\usepackage{sidecap}
\usepackage{xcolor}
\usepackage{subfig}
\usepackage{appendix}
\usepackage{rotating}
\bibpunct{(}{)}{;}{a}{}{,} 
\def \msun{\rm \, M_\odot}
\usepackage[switch]{lineno}

\usepackage{hyperref}
\usepackage{upgreek}

\hypersetup{
  colorlinks   = true, 
  urlcolor     = blue, 
  linkcolor    = red, 
  citecolor    = blue 
}

\def\cgs{erg~cm$^{-2}$~s$^{-1}$}

\begin{document}

\title{AGN feedback in an infant galaxy cluster:
the LOFAR-Chandra view of the giant FRII radio galaxy J103025+052430 at z=1.7}

\author{M. Brienza\inst{1,2,3}, 
R.~Gilli\inst{1},
I.~Prandoni\inst{3}, 
Q.~D'Amato\inst{4,3},
K.~Rajpurohit\inst{2,3,5},
F.~Calura\inst{1},
M.~Chiaberge\inst{6,7},
A.~Comastri\inst{1},
K.~Iwasawa\inst{8,9},
G.~Lanzuisi\inst{1},
E.~Liuzzo\inst{3},
S.~Marchesi\inst{1,10},
M.~Mignoli\inst{1},
G.~Miley\inst{11},
C.~Norman\inst{6,7},
A.~Peca\inst{12,1,2},
M.~Raciti\inst{13},
T.~Shimwell\inst{11,14},
P.~Tozzi\inst{15},
C.~Vignali\inst{1},
F.~Vitello\inst{3,13},
F.~Vito\inst{1}}

\institute{INAF – Osservatorio di Astrofisica e Scienza dello Spazio di Bologna, Via P. Gobetti 93/3, 40129 Bologna, Italy
\and 
Dipartimento di Fisica e Astronomia, Università di Bologna, via P. Gobetti 93/2, I-40129, Bologna, Italy
\and 
INAF - Istituto di Radioastronomia, Bologna Via Gobetti 101, I-40129 Bologna, Italy
\and 
SISSA, Via Bonomea 265, 34136 Trieste, Italy
\and
Harvard-Smithsonian Center for Astrophysics, 60 Garden Street, Cambridge, MA 02138, USA
\and
The William H. Miller III Department of Physics \& Astronomy, Johns Hopkins University, Baltimore, MD 21218, USA
\and
Space Telescope Science Institute for the European Space Agency, ESA Office, 3700 San Martin Drive, Baltimore, MD 21218, USA
\and
Institut de Ci\`encies del Cosmos (ICCUB), Universitat de Barcelona (IEEC-UB), Mart\'i i Franqu\`es, 1, 08028 Barcelona, Spain
\and 
ICREA, Pg. Llu\'is Companys 23, 08010 Barcelona, Spain
\and
Department of Physics and Astronomy, Clemson University, Kinard Lab of Physics, Clemson, SC 29634, USA
\and 
Leiden Observatory, Leiden University, PO Box 9513, 2300 RA Leiden, The Netherlands
\and
Department of Physics, University of Miami, Coral Gables, FL 33124, USA
\and
INAF - Osservatorio Astrofisico di Catania, Catania Via Santa Sofia 78, 95123 Catania, Italy,
\and
ASTRON, Netherlands Institute for Radio Astronomy, Oude Hoogeveensedijk 4, 7991 PD, Dwingeloo, The Netherlands
\and
INAF - Osservatorio Astrofisico di Arcetri, Largo E. Fermi, I-50122 Firenze, Italy}
   
\titlerunning{AGN feedback in an infant galaxy cluster}
\authorrunning{M. Brienza, R. Gilli et al.}

\date{Accepted ---; received ---; in original form \today}

\abstract 
{In the nearby universe jets from Active Galactic Nuclei (AGN) are observed to have a dramatic impact on their surrounding extragalactic environment. The effect of jets at high redshift (z>1.5) is instead much less constrained. However, studying their impact at the `cosmic noon', the epoch when both star formation and AGN activity peak, is crucial to fully understand galaxy evolution. 

Here we present a study of the giant ($\sim$750 kpc) radio galaxy 103025+052430 located at the centre of a protocluster at redshift z=1.7, with a focus on its interaction with the external medium. We present new LOFAR observations at 144 MHz, which we combine with VLA 1.4 GHz data and 0.5-7 keV Chandra archival data.
The new radio map at 144 MHz confirms that the source has a complex morphology, which can possibly fit the `hybrid morphology' radio galaxy classification. The large size of the source gave us the possibility to perform a resolved radio spectral index analysis, a very unique opportunity for a source at such high redshift. This reveals a tentative, unexpected, flattening of the radio spectral index at the edge of the backflow in the Western lobe, which might be indicating plasma compression. The spatial coincidence between this region and the thermal X-ray bubble C, suggests a causal connection between the two. 
Contrary to previous estimates for the bright X-ray component A, we find that inverse Compton scattering between the radio-emitting plasma of the Eastern lobe and the CMB photons can account for a large fraction ($\sim$ 45\%-80\%) of its total 0.5-7 keV measured flux. Finally, the X-ray bubble C, which is consistent with a thermal origin, is found to be significantly overpressurised with respect to the ambient medium. This suggests that it will tend to expand and release its energy in the surroundings, contributing to the overall intracluster medium heating. Overall, 103025+052430 gives us the chance to investigate the interaction between AGN jets and the surrounding medium in a system that is likely the predecessor of the rich galaxy clusters we all well know at z=0.

}

 \maketitle

\section{Introduction}
\label{sec:intro}

In the last decades increasing evidence of the impact of jetted Active Galactic Nuclei (AGN) on their surrounding medium has been collected in the local universe. On galactic scales (from a few kpc to a few tens of kpc) AGN jets have clearly shown to be able to compress, redistribute and even drive out of the host galaxy a considerable quantity of atomic, molecular and ionised gas (e.g. \citealp{miley1981, clark1997, wagner2012, morganti2013, santoro2018, mukherjee2018, zovaro2019, nesvadba2021, ruffa2022, murthy2022, capetti2022}). On larger, extra-galactic scales (hundreds of kpc), jets have shown as well to be able to displace the surrounding thermal intragroup/intracluster medium (IGrM/ICM), creating cavities in its X-ray distribution, but also to induce turbulence, shocks and sound waves, and to uplift metal-rich gas from the system's most central regions (see \citealp{mcnamara2007, fabian2012} for reviews).

The main reason why the action of AGN jets has attracted significant attention in the astrophysical community resides in its potential ability to regulate and even quench the formation of new stars in galaxies. This so-called negative AGN jet feedback is indeed strongly needed by simulations of galaxy evolution to prevent the spontaneous overcooling of gas in dark-matter halos and thus reproduce the observed galaxy mass function (e.g. \citealp{croton2006}).

Nevertheless, AGN jet feedback is a complex phenomenon and, under certain circumstances, the passage of the jet can play in favour of star formation triggering as well (positive feedback). Various examples of jet-induced star formation within the AGN host galaxy have been reported in literature (e.g. \citealp{vanbreugel1993, oosterloo2005, santoro2016, zinn2013}), finding also support in simulations (e.g. \citealp{mellema2002, gaibler2012, meenakshi2022}). Interestingly, there are a few cases where jets are reported to interact with and/or trigger star formation in external, neighbour galaxies (the Minkowski's object, \citealp{vanbreugel1985, croft2006, lacy2017, fragile2017}; 3C 441, \citealp{lacy1998}; HE0450-2958, \citealp{molnar2017} and DA 240, \citealp{chen2018}).

When moving to higher redshifts, the investigation of AGN jet feedback becomes more challenging mainly due to observational limitations, and thus its impact is less constrained (e.g. \citealp{hardcastle2020}). However, investigating the role jets play at the `cosmic noon' (z=1-3, \citealp{madau2014}), the epoch when both star formation and AGN activity peak, and beyond, is crucial to fully understand galaxy evolution.

Claims of positive, jet-induced feedback within the host of high redshift radio galaxies have been often made based on the so-called `alignment effect' \citep{rees1989}, that is the alignment of the optical emission along the direction of the radio axis. This co-spatiality is interpreted as evidence for massive star formation induced by shocks associated with the radio jet passage (e.g. \citealp{chambers1990, best1997, dey1997, lacy1999, bicknell2000}). Other findings for AGN jet feedback in high-z objects based on gas kinematics and outflows on galactic scales have also started to appear recently (e.g. \citealp{mukherjee2016, carniani2017, nesvadba2017}). Probing the effects of jets on the surrounding extragalactic environment remains instead a difficult task, which requires further efforts. 

It is also worth remembering that jetted AGN are predominantly associated to the most massive galaxies, which are located in rich environments. For this reason, high-redshift radio galaxies (z>1.5) are important tools to pinpointing protoclusters (e.g. \citealp{pentericci2000, miley2008, chiaberge2010, hatch2014, overzier2016, tozzi2022}) and to study AGN feeding, feedback, and mass assembly in these systems, whose fate is to eventually become the largest gravitationally bound structures of the local universe.

To date, the giant (>700 kpc) radio galaxy J103025+052430 located at the center of a protocluster at z=1.7 (\citealp{petric2003, nanni2018, gilli2019, damato2020, damato2021, damato2022}, is the first example at high redshift, where AGN jets seem to be able to trigger star formation in external galaxies. It therefore represents a very peculiar case of positive AGN jet feedback at cosmic noon, which deserves further investigation. Thanks to the availability of extraordinary deep (480 ks) \textit{Chandra} X-ray data \citep{nanni2018, gilli2019, nanni2020}, J103025+052430 is also one of the few radio galaxies around which extended X-ray emission associated with the radio jets has been detected at high redshift on scales of hundreds of kpc (see also \citealp{carilli2002, johnson2007, erlund2008, laskar2010}), ascribed to a combination of thermal (shocks) and non-thermal (inverse-Compton, IC, scattering) processes (see Sect. 2 for further discussion).

Here we present new dedicated radio observations of this system performed with the LOw-Frequency ARray (LOFAR, \citealp{vanhaarlem2013}) at 144\,MHz, aiming at recovering the oldest non-thermal particle population in the source. Using these data combined with recently published, deep 1.4\,GHz data \citep{damato2022} and the aforementioned X-ray data \citep{nanni2018, gilli2019}, we take new steps forward towards a full characterisation of the radio galaxy. In particular, we analyse its radio spectral properties and the interplay between its radio and X-ray emission, which we use to probe the interaction between the jet and the external gas and so, in turn, the jet feedback on the protocluster.

The paper is organised in the following way. In Sect.\,2 we give an overview of the radio galaxy and the system in which it is located.  In Sect.\,3  we describe the LOFAR 144-MHz observations and the related data reduction procedures. We also summarise the complementary data used in the subsequent analysis. In Sect.\,4 we describe and discuss the source properties at 144\,MHz, its resolved spectral properties in the frequency range 144-1400 MHz, and the connection between the observed radio and X-ray emission. In Sect. 5 our findings are summarised.

Hereafter, we assume a concordance $\rm{\Lambda}$CDM cosmology, with $\rm{\Omega_{m}}$ = 0.3, $\rm{\Omega_\Lambda}$ = 0.7 and H$_{0}$ = 70 km\,s$^{-1}$Mpc$^{-1}$, in agreement within the uncertainties with the Planck 2015 results \citep{planck2016}. In the adopted cosmology, the angular scale at the source redshift z=1.6987 is 8.5 kpc/arcsec. All coordinates are given in J2000. The spectral index $\alpha$ is defined as $S \propto \nu^{-\alpha}$, where $S$ is the flux density and $\nu$ is the frequency.

\section{The radio galaxy J103025+052430: an overview}

J103025+052430 is a powerful ($\rm P_{1400\,MHz}=4\times10^{26} \ W/Hz$) Fanaroff–Riley II (FRII, \citealp{fanaroff1974}) radio galaxy, associated with a massive galaxy ($M_*\sim3\times10^{11} \msun$) located at the center of a galaxy protocluster at redshift z=1.6987 (\citep{petric2003, nanni2018, gilli2019, damato2020, damato2021, damato2022}. It is powered by a supermassive black hole with mass $\rm <3.2\times10^7 \ \msun$ as estimated by \cite{gilli2019} based on the bolometric luminosity. With its total projected extension ($\geq$700 kpc), it is one of the few giant radio sources at z>1.5 studied to date (see e.g. \citealp{erlund2008, kuzmicz2021} for other examples), providing a unique opportunity to study feedback from AGN jets at high redshift, on large (hundreds of kpc) scales.

\begin{figure*}[!htp]
\centering
\includegraphics[width=0.8\textwidth]{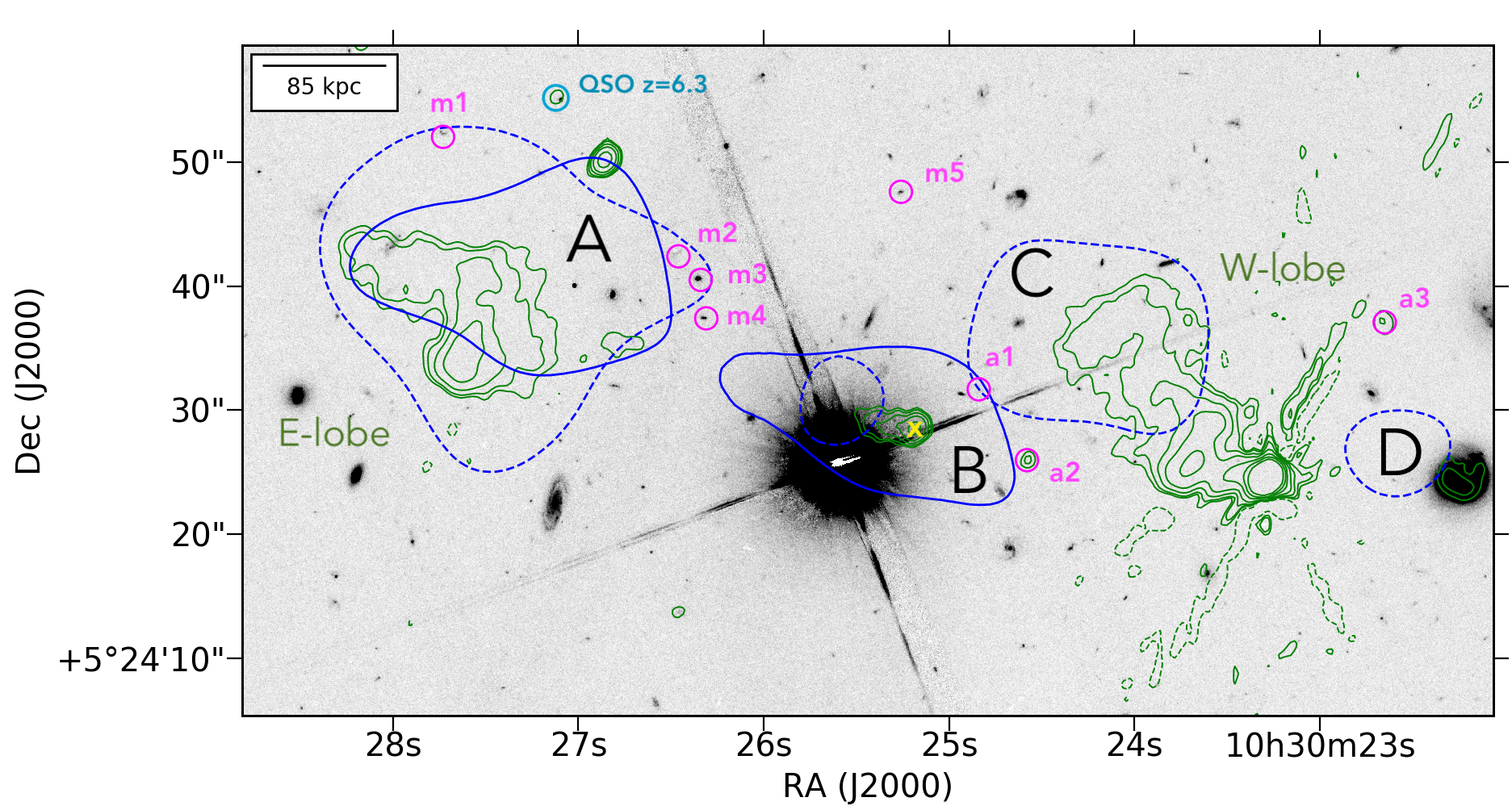}
\caption{\small{Image of the FRII radio galaxy J103025+052430. In background the Hubble Space Telescope F160W-filter image is shown \citep{damato2020}. Radio JVLA contours (starting from 3$\rm \sigma$) at 1400 MHz with 1.5$\times$1.2 arcsec beam \citep{damato2022} are overlaid in green. The X-ray \textit{Chandra} diffuse emission above 2.5$\rm \sigma$ in the soft band (0.5-2 keV) and hard band (2-7 keV) is shown with blue dashed contours and blue solid contours, respectively \citep{nanni2018, gilli2019}. Black letters A,B,C,D mark the main X-ray components as defined in \cite{gilli2019}. Magenta circles mark the MUSE (m1-m5, \citealp{gilli2019}) and ALMA (a1-a3, \citealp{damato2020}) galaxies, which are part of the protocluster at z=1.7 presented in \cite{gilli2019}. The SDSSJ1030+0524 quasar at z=6.3 is marked with a cyan circle. The FRII radio galaxy host is marked with a yellow cross. Note that two galaxy members of the protocluster are located outside the displayed field of view: m6, located at $\sim$25 arcsec North of m2, and one source detected by LUCI located at $\sim$1.5 arcmin southeast of the FRII radio galaxy host.}}
\label{fig:overview}
\end{figure*}

The source was first serendipitously discovered and classified as an FRII by \cite{petric2003}, who used the Very Large Array (VLA) at 1.4\,GHz to observe the field around the bright optically-selected quasar SDSSJ1030+0524 (RA = 10h30m27s, Dec = +05d24'55", \citealp{fan2001}) at z=6.3. New, deeper (2.5~$\mu$Jy/beam with beam $\sim$1 arcsec) observations at 1.4\,GHz performed with the Karl G. Jansky Very Large Array (JVLA) have more recently been collected and analysed (see green contours in Fig. \ref{fig:overview}), providing a more detailed view of the radio continuum and polarisation properties of the source \citep{damato2021, damato2022}. In particular, these data have revealed more extended emission than previously known associated with both the Eastern and Western lobe with peculiar, bent morphologies and a fractional polarisation of 10\%-20\% in correspondence of the brightest regions. Interestingly, along the jet/lobe bending, the fractional polarisation increases and the magnetic field is aligned, suggesting possible compression of the non-thermal plasma.

The SDSSJ1030+0524 quasar field is very well studied thanks to a rich multi-wavelength observational campaign performed by our group. Results and data products can be found in the project webpage\footnote{\url{http://j1030-field.oas.inaf.it/data.html}}. This also allows for a detailed investigation of the FRII. In particular, \cite{nanni2018} detected diffuse, extended X-ray emission using \textit{Chandra} observations in the band $0.5{-}7$ keV, mostly coinciding with its eastern lobe and central nucleus (see blue and red contours in Fig. \ref{fig:overview}). 

Later, using spectroscopic data collected with the Multi Unit Spectroscopic Explorer (MUSE) at the Very Large Telescope (VLT) and the Large Binocular Telescope (LBT) with the LBT Utility Camera in the Infrared (LUCI), \cite{gilli2019} measured the redshift of the FRII and identified a galaxy overdensity composed of seven star-forming (star formation rate, SFR$\rm \sim8-60 \ \msun/yr$) galaxies around it (five of them are visible in Fig. \ref{fig:overview} and labelled as m1-m5).

Based on Atacama Large (sub-)Millimeter Array (ALMA) observations, three more gas-rich members of the system were later identified by \cite{damato2020}, with star formation rates in the range SFR$\rm \sim5-100 \ \msun/yr$ (labelled as a1-a3 in Fig. \ref{fig:overview}). 

The same FRII host is a highly star forming galaxy with SFR $\rm 570 \ \msun/yr$ based on galaxy SED fitting \citep{damato2021}. All member galaxies are distributed within a projected distance of 1.15 Mpc from the FRII host and span a redshift range of $\rm \Delta$z=0.012.
The system is likely still not virialised and based on the overdensity volume probed has a total mass of $\geq 3\times10^{13} \msun$ \citep{gilli2019, damato2020}. Likely, it represents the early assembly phase of what will become a massive galaxy cluster (M$\rm_{sys}$>$\rm10^{14} \msun$) in the z=0 universe \citep{damato2020}. In this view, it is also likely that the FRII host will evolve into the brightest cluster galaxy (BCG). 

In the optical band and based on the NII/H$\rm\alpha$ ratio and narrow width of H$\rm\alpha$, the FRII host is classified as a classical type-2 AGN \citep{gilli2019}. The nuclear X-ray emission is also consistent with a heavily obscured (Compton thick) AGN, with a derived column density of $\rm N_{H,X}\sim10^{24} \ cm^{-2}$ \citep{gilli2019}. Its bolometric luminosity equal to $\rm L_{bol}=4\times10^{45}$ erg/s (as derived from the absorption-corrected X-ray luminosity equal to $\rm L_{2-10keV}=1.3\times10^{44}$ erg/s, assuming a correction factor of 30 following \citealp{marconi2004}), further suggests a classification as a quasar.  Recent ALMA observations by \cite{damato2020} showed that the FRII host is surrounded by an extended ($\approx 20$ kpc) and massive ($\rm M_{H2}=2\times10^{11} \msun$) molecular gas reservoir, which may significantly contribute to the observed X-ray column density.

Using the \textit{Chandra} observations presented by \cite{nanni2018}, the deepest so far for a distant FRII within a galaxy overdensity (480\,ks), \cite{gilli2019} suggested that the observed extended X-ray emission is likely due to a combination of non-thermal emission produced by IC scattering between the photons of the cosmic microwave background (CMB) and the relativistic electrons in the radio lobe, and thermal emission produced by an expanding bubble of gas that is shock-heated by the FRII jet.

Interestingly, four out of the six MUSE galaxies in the overdensity are distributed in an arc-like shape right at the edge of the diffuse X-ray emission around the eastern lobe of the FRII (see galaxies m1-m4 in Fig. \ref{fig:overview}). The measured specific star formation rates (sSFR) in these galaxies are a factor from two to five higher than those measured in the other protocluster members and those of typical main sequence galaxies of equal stellar mass and redshift \citep{gilli2019}. This striking spatial coincidence and enhanced sSFR were interpreted as unique evidence of positive AGN feedback on multiple galaxies at high redshift, i.e. star formation promoted by the expanding lobes, making it a unique object.

\section{Data}
\label{sec:data}

\subsection{LOFAR observations at 144 MHz}
\label{sec:lofar}

The target was observed with the LOFAR High-Band Antennas in the frequency range 120-168\,MHz for a total observing time of 16 hours (see Table \ref{tab:obs} for the observation setup). To obtain the best possible uv-coverage given the low declination of the target, the total observing time was split into 4 hours observing blocks. The observation setup was created following the standard LOFAR Two-meter Sky Survey (LoTSS; \citealp{shimwell2019}) strategy. Both Dutch and international stations were included in the observations but the latter were not analysed here and will be presented in a future work. All four polarisations were recorded (XX, XY, YX, YY), the correlator time was set to 1s and the frequency resolution to 12.2\,kHz. 3C 295 was observed for 10 min at the beginning of each observing run and used as primary calibrator. 

\begin{table}[!htp]

        \small
 \caption{\small{LOFAR observation details.}}
        \centering
                \begin{tabular}{l l}
                \hline
                \hline
         
Project code & LC12\_012 (PI R. Gilli)\\
Observation dates & 07-15/07/2019, 04/10/2019, 15/11/2019\\
Telescope configuration & HBA Dual inner \\
Central Frequency   & 144 MHz\\
Bandwidth & 47 MHz\\
Channel width  & 12.2 kHz \\
Integration time & 1s\\
Observation duration & 4$\times$4h\\
Polarisation & Full Stokes\\
Primary calibrator & 3C 295\\

                \hline
                \hline  
                \end{tabular}
     \label{tab:obs}
\end{table}

Before being stored in the LOFAR long-term archive, the data were pre-processed using the observatory pipeline \citep{heald2010}. This performed automatic flagging of radio frequency interference using the {\tt AOFlagger} \citep{offringa2012} and averaged the data by a factor of 4 in frequency.

\begin{figure*}[!htp]
\centering
\includegraphics[width=0.75\textwidth]{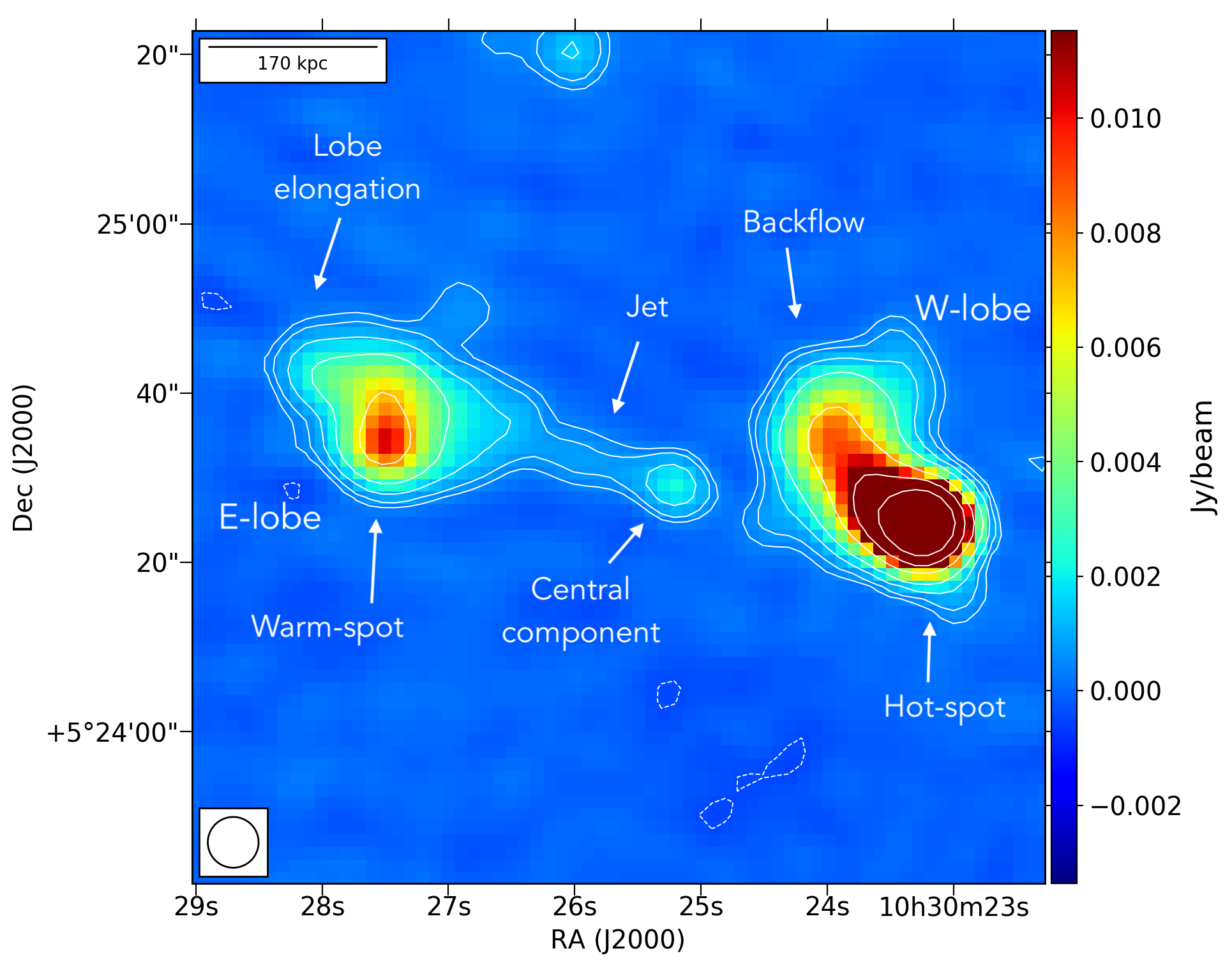}
\caption{\small{LOFAR image of the radio galaxy J103025+052430 located at redshift z=1.7 at a central frequency of 144\,MHz. Labels mark the most significant morphological features in the source. The beam equal to 6 arcsec $\times$ 6 arcsec is shown in the bottom left corner. Contours are drawn at [-3, 3, 5, 10, 20, 50, 100, 200]$\rm \times \sigma$, with $\rm \sigma$=0.14 mJy/beam. }}
\label{fig:lofar}
\end{figure*}

Using the {\tt PreFactor} pipeline\footnote{\url{https://github.com/lofar-astron/prefactor}} as described by \cite{vanweeren2016} and \cite{williams2016}, direction-independent calibration was performed, which corrects for ionospheric Faraday rotation, offsets between XX and YY phases and clock offsets (see \citealt{degasperin2019}).
Ionospheric distortions and errors in the beam model were then corrected using the  DDF-pipeline\footnote{\url{https://github.com/mhardcastle/ddf-pipeline}}, which performs various direction-dependent self-calibration cycles (\citealp{tasse2021} and \citealp{shimwell2019}) using the software {\tt kMS} (\citealt{tasse2014} and \citealt{smirnov2015}) and the imager {\tt DDFacet} (\citealt{tasse2018}). The calibration in the direction of the target was further improved by applying the procedure presented in \cite{vanweeren2020}. This consists of subtracting from the visibilities all sources outside a square region centred on the target (here with side equal to 24 arcmin), and performing a final direction-independent self-calibration round.

The final image deconvolution was performed with {\tt WSClean} (version 2.8, \citealp{offringa2014}) using a {\tt Briggs} weighting scheme with ${\tt robust}=-0.5$ (which optimises spatial resolution and noise pattern), multiscale cleaning and applying an inner uv-cut at 40k$\rm \lambda$. The final image (presented in Fig. \ref{fig:lofar}) was restored with a beam of 6 arcsec $\times$ 6 arcsec (51 kpc $\times$ 51 kpc) and has an rms of 0.14 mJy $\rm beam^{-1}$. 

A second image was also produced using the same aforementioned parameters, but excluding all baselines shorter than 1450$\rm \lambda$. This cut corresponds to the shortest well-sampled baseline of the JVLA dataset at 1400\,MHz (see Sect. \ref{sec:otherdata}) and is applied to recover at both frequencies the flux density on the same spatial scales and to perform a reliable spectral analysis (see Sect. \ref{sec:spec}). 

The flux density scale of the final image was checked using as a reference the ten brightest sources in the field. The flux densities of these sources from all publicly available surveys were used to extrapolate the expected flux densities at the frequencies of interest and these were compared with the measured values. Following this procedure, we did not find any systematic offset in the flux density scale of the image within a 1$\rm sigma$ uncertainty. We assume a conservative flux density scale calibration uncertainty equal to $\Delta S_c$=20\% following \cite{shimwell2019, shimwell2022}. The flux density of the target is consistent with what measured in the TIFR GMRT Sky Survey-Alternative Data Release (TGSS-ADR1; \citealp{intema2017}) within uncertainties.

\subsection{Other data}
\label{sec:otherdata}

Beside the new LOFAR data at 144\,MHz described in Sect.\,\ref{sec:lofar}, for the analysis of this work, we also used the JVLA dataset at 1400\,MHz, recently presented in \citep{damato2022}. This was obtained using A-array observations for a total observing time of $\sim$30 hours. For a full description of the data reduction procedures, we defer the reader to \citep{damato2022}. Here we re-imaged the calibrated data to match the properties of the LOFAR image and perform a resolved spectral analysis of the source in the frequency range 144-1400\,MHz (see Sect.\,\ref{sec:spec}). To create the image, we used {\tt WSClean} and the following parameters: {\tt Briggs} weighting scheme with ${\tt robust}=0$, gaussian taper of 5 arcsec and a final restoring beam equal to 6 arcsec $\times$ 6 arcsec. The final image has an rms of 0.025 mJy $\rm beam^{-1}$. A flux density scale uncertainty equal to $\Delta S_c$=5\% is considered following \citet{perley2013}.

The LOFAR and VLA images were spatially aligned to correct for spatial shifts introduced by the self-calibration. To do this, we used a bright point source in the field (RA=10h30m16s and DEC=+05d23m03s). The source was fitted with a 2D-Gaussian function and its central pixel position was used as a reference for alignment. The procedure was carried out using the tasks {\tt imfit}, {\tt imhead} and {\tt imregrid} in the Common Astronomy Software Applications ({\tt CASA}, \citealp{mcmullin2007}) package. After this procedure, the images have a residual offset <0.1 pixels, which is sufficiently accurate for our analysis. 

Finally, we use the \textit{Chandra} images in the energy range 0.5-7 keV, presented in \cite{nanni2018} and \cite{gilli2019}, for a comparison between the radio and X-ray emission.

\section{Analysis and results}

\subsection{Radio source properties}
\label{sec:morpho}

In Fig.\,\ref{fig:lofar}, we present our new LOFAR image of the radio galaxy J103025+052430 at 144\,MHz. The most relevant features are labelled in the figure. The source features two lobes and a central unresolved component, consistent with what was already observed at 1400\,MHz \citep{petric2003, damato2022}. Moreover, a large-scale jet-like structure connecting the Eastern lobe to the AGN nucleus is now clearly detected.

The Western lobe shows a clear FRII-like structure, including a compact hot-spot and a backflow bending towards North-West. As already reported by \citep{damato2022}, the morphology of the Eastern lobe is more unusual. It shows a surface brightness (SB) enhancement towards the southern edge, which does not satisfy the classical definition of hotspot (i.e. SB$\geq$0.6 mJy/arcsec$\rm^2$ and SB contrast with respect to the entire lobe $\geq$4, \citealp{deruiter1990}) and is therefore referred to as \textit{warm spot}. Moreover, in North-East direction with respect to the warm spot the Eastern lobe shows a further elongation (see Fig.\,\ref{fig:lofar}). Both the jet/counterjet ratio and the polarisation analysis of the source suggest that the jet terminates in the warm spot \citep{damato2021, damato2022}, as detailed below. The polarisation image of the source shows two patches of emission in correspondence of the warm-spot, similar to what is usually observed in the hotspots of FRII radio galaxies and suggesting this is the region where the jet impacts the surrounding medium and gets compressed. Moreover, the magnetic field orientation experiences a 90-degree rotation moving from the warm-spot peak towards the center of the lobe to the north, which might be indicative of the jets bending towards north after impacting the ICM. Furthermore, the radio galaxy inclination angle estimated from the jet versus counter-jet base flux density ratio \citep{gilli2019, damato2022} is in better agreement with that estimated from the jet/counter-jet  length ratio, when it is assumed that the jet ends at the warm spot, rather than at the end of the northern elongated region.

The origin of the complex morphology of the Eastern lobe is unclear, despite not unusual in giant radio galaxies (e.g. \citealp{bruni2020, cantwell2020, carilli2022}. It could be related to the jet being precessing \citep{donohoe2016, horton2020}, being bent with respect to the observer's line of sight \citep{harwood2020}, or to the jet flow being deviated by a patchy external environment. The latter option would be particularly consistent with an dishomogeneous ambient medium, as expected in a protocluster environment, and could further justify the asymmetry with the Western lobe-core axis \citep{pirya2012, cantwell2020}. 

Overall, based on its morphology the radio galaxy could be classified as a HYbrid MOrphology Radio Source (Hymor, \citealp{gopalkrishna2000, harwood2020}), a rare class of radio galaxies in which a different FRI/FRII morphology is observed for each of the two lobes. 

The central component of the radio galaxy at 144 MHz is unresolved, implying an upper limit to its physical extension of $\sim$ 50\,kpc. As we can see from observations at 1400\,MHz with higher resolution (Fig. \ref{fig:overview}), this region includes both the actual nuclear emission and the base of the Eastern jet.

Using the 3$\sigma$ contours as a reference, the total projected extension of the source in the LOFAR image is $\sim$90 arcsec, which corresponds to $\sim$750 kpc, consistent with previous estimates made using the 1.4 GHz image \citep{damato2022}. This places the source within the giant radio galaxy population (GRG, \citealp{kuzmicz2021}). Based on considerations on the jet-counterjet flux density ratio and length ratio, \citep{damato2022} computed that the source inclination with respect to line of sight is equal to $\sim$80 deg, with the Eastern jet approaching the observer. By accounting for the source inclination, we obtain that the source intrinsic extension must be 750$\rm / cos\theta\approx$800 kpc. 

We stress that, while recent observational efforts are significantly increasing the detection number of giant radio galaxies and giant radio quasars (GRQs) (e.g. \citealp{kuzmicz2018, dabhade2020a, dabhade2020b, bruni2020, kuzmicz2021, delhaize2021, heinz2021, simonte2022, mahato2022}, \citealp{oei2022}), the number of giant radio sources identified at redshift $z>1.5$ remains to date limited (10/239 in the sample of \citealp{dabhade2020b}; 33/272 in the sample of \citealp{kuzmicz2021}, 3/178 in the sample of \citealp{heinz2021}, 7/74 in the sample of \citealp{simonte2022}), possibly due to observational limits.

\begin{figure*}[!htp]
\centering
\minipage{0.5\textwidth}
\centering
\includegraphics[width=1\textwidth]{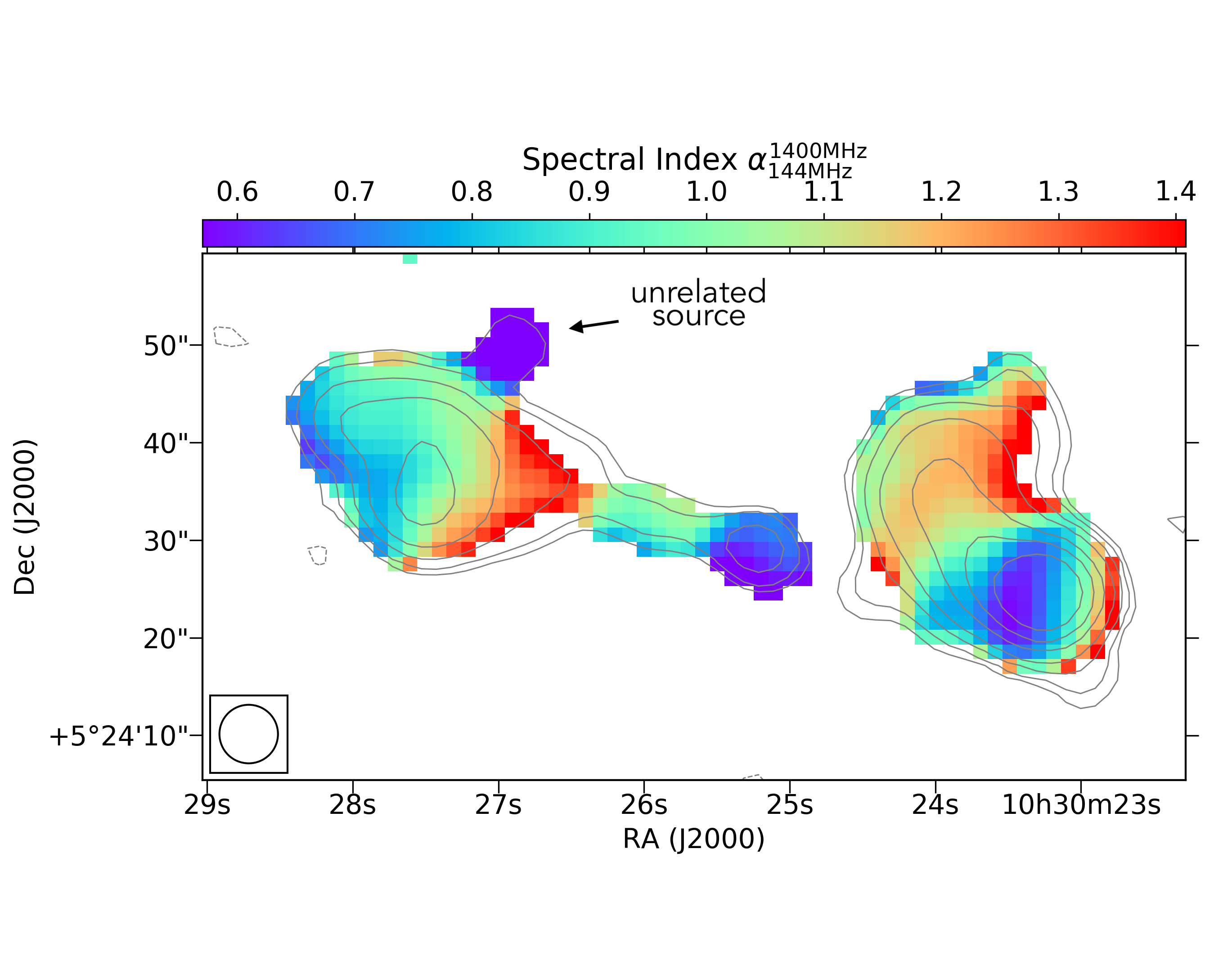}
\endminipage\hfill
\centering
\minipage{0.50\textwidth}
\centering
\includegraphics[width=1.02\textwidth]{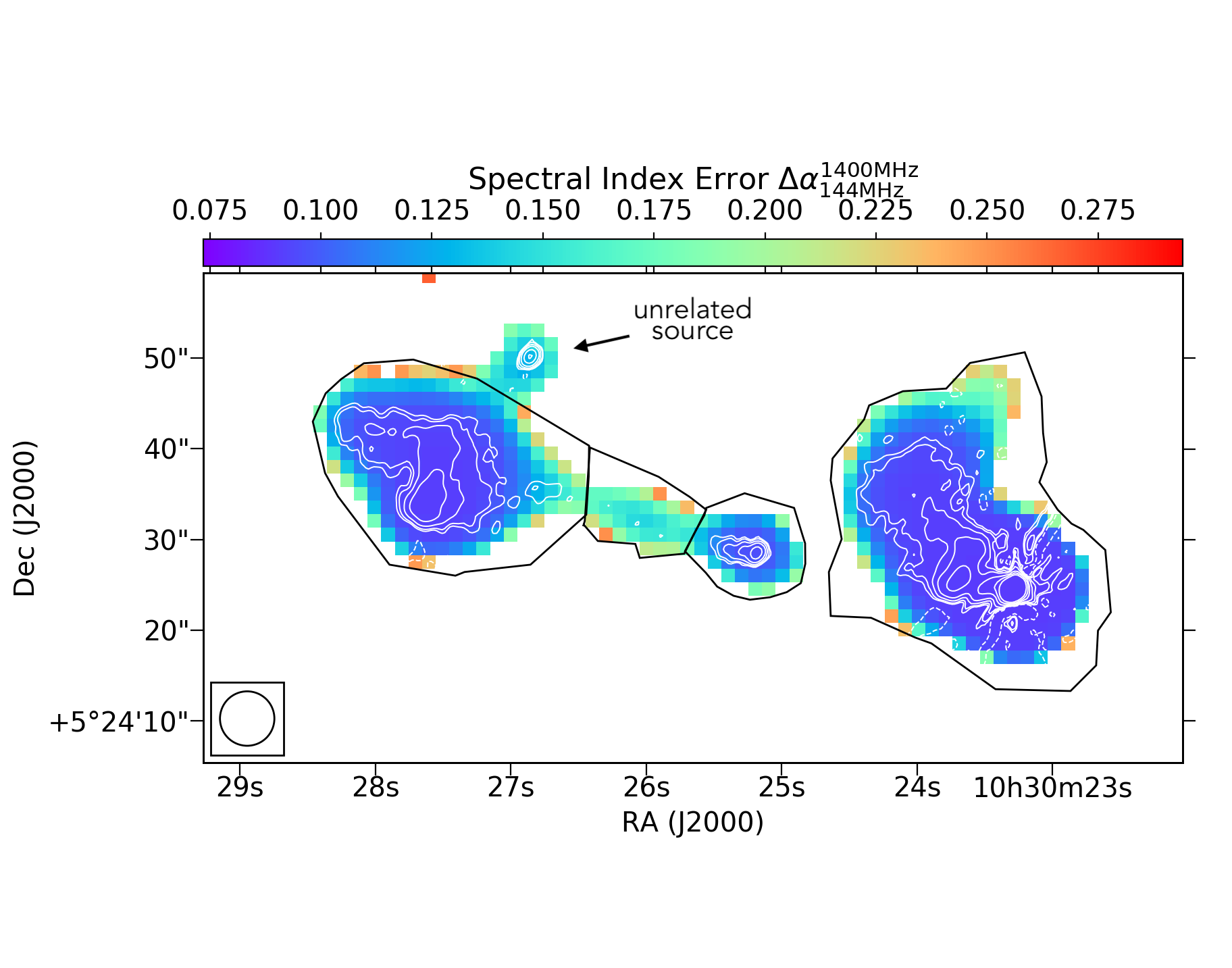}
\endminipage\hfill
\caption{\small{Radio spectral index map of J103025+052430 in the range 144-1400 MHz with 6-arcsec resolution (left panel) and respective spectral index uncertainty map (right panel). The spectral index values across the radio galaxy vary in the range [0.46-1.78] but the colorscale has been compressed for visualisation purposes. Only pixels above $3\sigma$ in both images have been used. The beam size is shown in the bottom-left corner. Contours in the left panel represent the 144-MHz emission as shown in Fig. \ref{fig:lofar} (left panel). White contours in the right panel represent the JVLA 1400-MHz emission at 1-arcsec resolution presented in \citep{damato2022}. with levels, equal to [-3,3,5,10,20,50,100,200]$\rm \times \sigma$, with $\rm \sigma$=0.004 mJy/beam. Black regions in the right panel represent the regions used to measure the flux densities listed in Table \ref{tab:flux}. The spectral index distribution of the source is overall consistent with what is typically observed in FRII radio galaxies. The compact source at RA=10h30'27" DEC=5d24m50s is an unrelated source.}}
\label{fig:spec}
\end{figure*}

The total flux density of J103025+052430 at 144\,MHz and 1400\,MHz, as well as the flux densities of the individual morphological components, are listed in Table\,\ref{tab:flux}, together with the respective luminosities. These were measured from the 6-arcsec resolution maps at both frequencies for consistency and using the black regions shown in Fig. \ref{fig:spec}, right panel, which were drawn following the 3$\rm \sigma$ contours at 144 MHz. The total uncertainty ($\Delta S$) on the flux densities ($S$) were computed by combining in quadrature the flux density scale calibration uncertainty ($\Delta S_c$) and the image rms noise ($\sigma$) multiplied by the flux density integration area in beam units ($A_{int}$):

\begin{equation}
\label{eq:fluxerr}
    \Delta S = \sqrt{(S\cdot\Delta S_c)^2+(\sigma \cdot A_{\rm int})^2}
\end{equation}

\begin{table*}[!htp]

 \caption{\small{Source properties as measured in the 6-arcsec images.}}
        \centering
                \begin{tabular}{c c c c c c}
                \hline
                \hline
         Region & $\rm S_{144\,MHz}$ & $\rm S_{1400\,MHz}$ & $\rm \alpha_{144\,MHz}^{1400\,MHz}$ & $P_{144\,\rm MHz}$ & $P_{1400\,\rm MHz}$ \\
                         
         & [mJy] & [mJy] & & [$\rm \times 10^{25}$ W/Hz] & [$\rm \times 10^{24}$ W/Hz] \\
         \hline 
         
         Total & 180$\pm$40 & 26.8$\pm$1.5 & 0.84$\pm$0.09 & 309$\pm$12 & 457$\pm$4 \\
         
         Western lobe & 140$\pm$30 & 22$\pm$1 & 0.81$\pm$0.09 &  237$\pm$9& 373$\pm$4 \\
         
         Eastern lobe & 35$\pm$7 & 3.6$\pm$0.2 & 0.99$\pm$0.09 & 68$\pm$2 & 72.0$\pm$0.6    \\
         
         Central component & 2.4$\pm$0.6 & 0.57$\pm$0.06 & 0.62$\pm$0.11 & 3.3$\pm$0.2 & 7.9$\pm$0.3   \\
        
         Eastern jet & 1.5$\pm$0.5 & 0.15$\pm$0.05 & 1.0$\pm$0.2 & 3.1$\pm$0.1 & 2.9$\pm$0.1\\
                \hline
                \hline  
                \end{tabular}
     \label{tab:flux}
\end{table*}

The integrated spectral indices and k-corrected radio powers are also presented in Table\,\ref{tab:flux}. The spectral index uncertainties were estimated with the following formula based on the propagation of error:

\begin{equation}
\rm \Delta\alpha = \frac{1}{ln\frac{144}{1400}}\sqrt{\left(\frac{\Delta S_{144MHz}}{S_{144MHz}}\right)^2+\left(\frac{\Delta S_{1400MHz}}{S_{1400MHz}}\right)^2}
\label{eq:spixerr}
\end{equation}

where $\rm {S_{144MHz}}$ and $\rm {S_{1400MHz}}$ are the flux density values at the respective frequencies in MHz and $\rm \Delta S_{144MHz}$ and $\rm \Delta S_{1400MHz}$ their corresponding uncertainty. 

With a power of $P_{144\,\rm MHz}\sim3\times10^{27}$ W/Hz, J103025+052430 appears to be among the 30 most powerful giant radio sources (with $ P_{144\,\rm MHz}>10^{27}$ W/Hz) in the sample of 239 GRGs and GRQs (optically selected, see \citealp{duncan2019}) extracted by \cite{dabhade2020b} from LoTSS.

The integrated spectral index of the source is $\rm \alpha_{144\,MHz}^{1400\, MHz}=0.84\pm0.09$ and lies well within the observed integrated spectral index distribution derived for the aforementioned sample: the full range is equal to $\rm \alpha_{144\,MHz}^{1400\,MHz}$=[0.4,1.5], and the median values are equal to $\rm \bar\alpha_{144\,MHz}^{1400\,MHz}=0.79$ and $\rm \bar\alpha_{144\,MHz}^{1400\,MHz}=0.76$ for GRGs and GRQs, respectively. These values are not different from those observed in active radio galaxies of smaller size (e.g. \citealp{parma1999, mullin2008}), thus confirming that despite the large size, giant radio sources are not remnant (inactive) sources and their integrated spectral index is dominated by freshly accelerated particles. The measured spectral index of J103025+052430 is also consistent with values observed for the ten sources at highest redshift ($1.5<z<2.3$) in the same sample.

\begin{figure}[!htp]
\centering
    \includegraphics[width=0.43\textwidth]{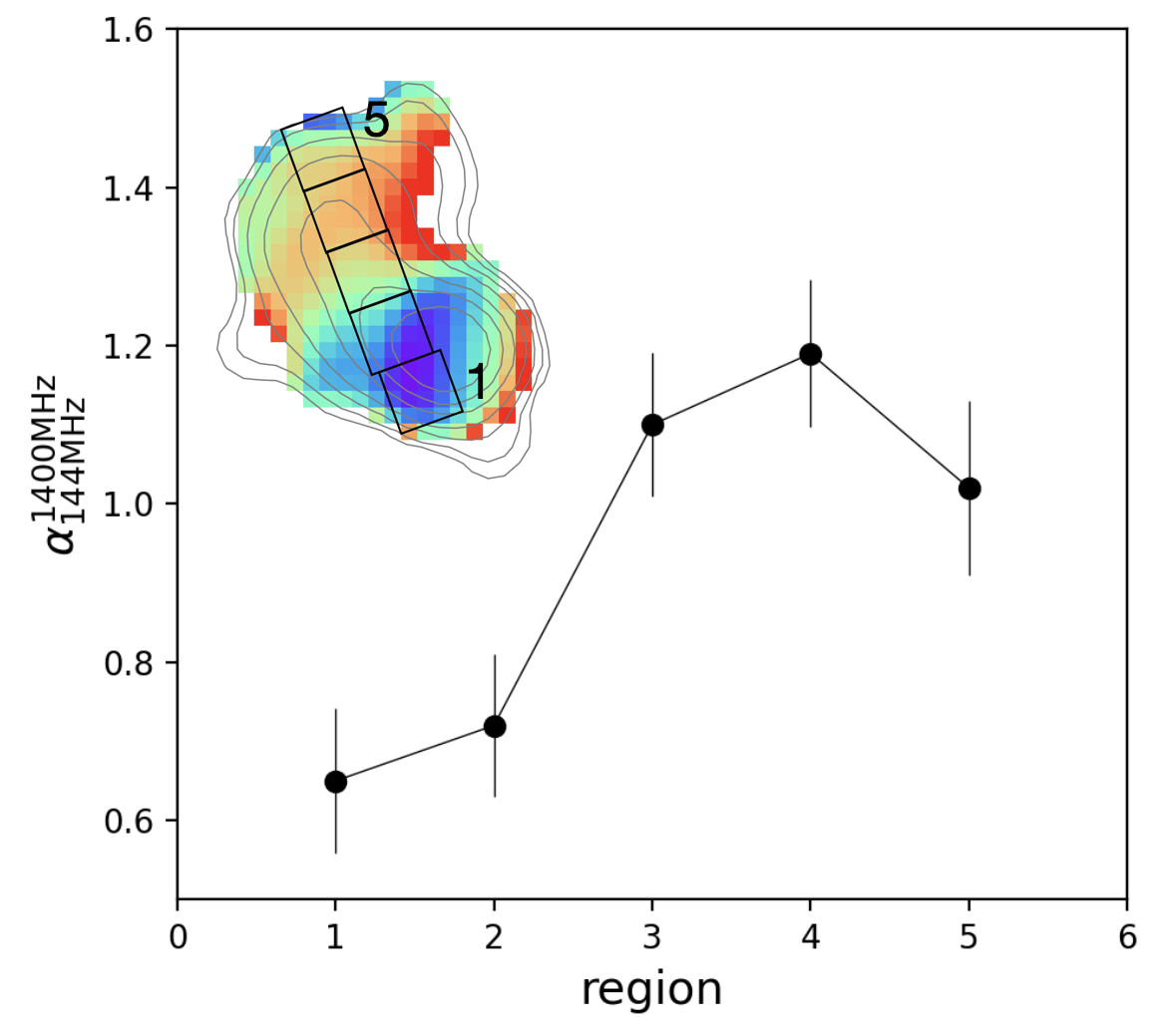}

\caption{\small{Spectral index profile in the frequency range 144-1400 MHz across the Western lobe of J103025+052430. The regions used to extract the spectral profile are shown as black boxes overlaid on the lobe in the top-left corner and have sizes equal to the beam dimension (6 arcsec=50 kpc). In the outermost region of the backflow, the spectral index shows a tentative flattening as discussed in Sect. \ref{sec:spec}.}}
\label{fig:specprofile}
\end{figure}

\subsection{Radio spectral index distribution}
\label{sec:spec}

To investigate the spectral index variations across the source, we created the spectral index map presented in Fig.\,\ref{fig:spec}, left panel, by combining the matched images at 144\,MHz and 1400\,MHz described in Sect.\,\ref{sec:data}. Only pixels with surface brightness values above 3$\sigma$ in both maps were included. The spectral index uncertainty map is shown in Fig. \ref{fig:spec}, right panel, and was obtained using Eq.\,\ref{eq:spixerr}, where ${S_{144}}$ and ${S_{1400}}$ correspond this time to the surface brightness values of each pixel at the respective frequencies, and $\Delta S$ is computed with Eq. \ref{eq:fluxerr} using $A_{\rm int}$=1.

The spectral index values across the source vary in the range [0.46-1.78]. Overall, the observed spectral index distribution supports the FRII classification made previously based on the source morphology only. In the lobes of these sources, in fact, the flattest spectral index values are observed at the lobe edges, where the actual particle acceleration occurs. The compact component at the center shows a flatter spectral index with a mean value of $\alpha_{144\,\rm MHz}^{1400\,\rm MHz}=0.6\pm0.1$, consistent with being produced by a superposition of nuclear activity and the base of the jet (e.g. \citealp{orru2010, harwood2016}).

In the Eastern lobe the mean spectral index is $\rm \alpha_{144\,\rm MHz}^{1400\,\rm MHz}=1.0\pm0.1$. Interestingly, from the spectral index map (Fig.\,\ref{fig:spec}) we can see that the flattest spectral index is not observed exactly in coincidence with the warm spot, which was previously claimed to represent the jet termination region. Instead the region with flatter spectral index (with values down to $\rm \alpha_{144\,\rm MHz}^{1400\,\rm MHz}=0.85\pm0.1$) is located along the entire edge of the Eastern lobe. No special spectral index trend is instead observed in correspondence of the polarisation enhancement and magnetic field alignment reported by \cite{damato2021} at the northern edge of the lobe, which were suggested to be consistent with the lobe being compressed. A flatter spectral index at this location would have been expected in such case.

The spectral index in the Western hot-spot has a minimum value of $\rm \alpha_{144\,\rm MHz}^{1400\,\rm MHz}=0.6\pm0.1$, consistent with what is observed in other similar sources (e.g. \citealp{vaddi2019}). As shown in Fig.\,\ref{fig:spec}, when moving away from the hotspot and across the backflow towards North, the spectral index steepens up to values of $\rm \alpha_{144\,\rm MHz}^{1400\,\rm MHz}=1.3\pm0.1$, as expected for ageing plasma \citep{pacholczyk1970}. Interestingly, moving towards the North-Eastern edge of the Western lobe, there is a tentative indication that the spectral index flattens again.

To investigate this trend further, we created a spectral index profile using five square regions distributed along the lobe, having side equal to 6 arcsec ($\sim$50 kpc), corresponding to the beam size (see Fig. \ref{fig:specprofile}, left panel). We note that the result is the same if we slightly change the orientation of the boxes. From the profile, we can see that the spectral index reaches a maximum value of $\rm \alpha_{144\,MHz}^{1400\,MHz}=1.2\pm0.1$ in region 4 and then decreases to a value of $\rm \alpha_{144\,MHz}^{1400\,MHz}=1.0\pm0.1$ in region 5.

This trend might be suggesting that the backflow is compressed by the interaction with the surrounding medium. Indeed, compression causes an enhancement of the number density of relativistic electrons and of the magnetic field. This translates into an increase of brightness, as well as a shift of the spectrum towards higher frequencies. This happens because the bulk of the synchrotron radiation released by each particle of energy $E=\gamma m_e c^2$ is emitted at the critical frequency $ \nu_c$:

\begin{equation}
    \nu_c=\frac{\gamma^2 e B}{2\pi m_e c},
\end{equation}
where $\gamma$ is the Lorentz factor, $m_e$ the electron mass, $c$ the speed of light, $e$ is the electron charge and $B$ is the magnetic field. If the magnetic field increases in the compressed plasma, the critical frequency of each particle will also increase, causing an overall shift of the entire spectrum towards higher frequencies. 

\begin{figure*}[!htp]
\centering
\includegraphics[width=0.7\textwidth]{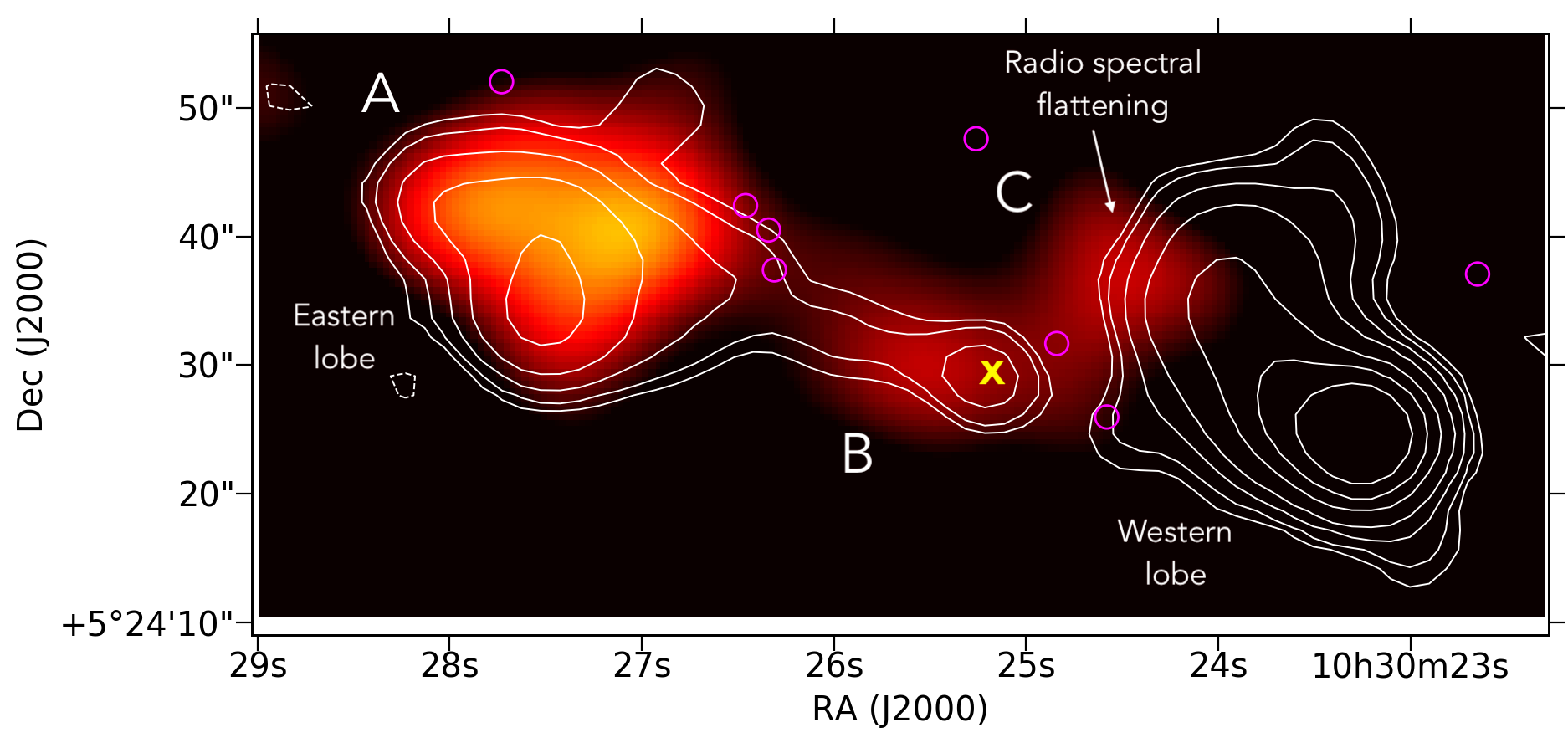}
\caption{\small{Diffuse X-ray emission (components A, B and C) as seen by \textit{Chandra} in the full 0.5-7 keV band with LOFAR 144-MHz contours overlaid in white (same as Fig. \ref{fig:lofar}). The  protocluster galaxy members as reported in Fig. \ref{fig:overview} are marked with magenta circles and the radio galaxy host is marked with a yellow cross. }}
\label{fig:chandra}
\end{figure*}

As shown in Fig.\,\ref{fig:chandra}, the region where the flattening is observed coincides with the presence of diffuse X-ray emission labelled as component C (see Sect. \ref{sec:ccomp}). The measured trend of the radio spectral index might indicate that, at that location, the radio-emitting plasma is indeed interacting with and compressing the external thermal medium, enhancing its X-ray brightness. Unfortunately, due to the low radio brightness of the backflow emission at 1.4 GHz, \cite{damato2021} could not detect any polarisation signal, which could have helped to investigate the compression scenario. Future multi-frequency radio observations at higher resolution are needed to confirm this trend.

Using the broadband radio astronomy tools soft-ware package (BRATS, Harwood et al. 2013) we modelled that it takes 8 Myr for the plasma to get a spectral index equal to $\rm \alpha^{1400MHz}_{144MHz}$=1.8, which is the steepest spectral index value observed in the radio galaxy. For this first order age derivation we used a Jaffe-Perola radiative model \citep{jaffe1973} and assumed an injection index of $\rm \alpha_{inj}$=0.6 (assumed to be the same as the spectral index observed in the hot-spot, where particle acceleration occurs) and a magnetic field equal to 1.3-2 $\mu$G (see Sect. \ref{sec:acomp} for a full discussion on the magnetic field derivation). We note, however, that the above estimate is almost independent from the assumed magnetic field as the particle energy losses at this redshift are dominated by IC losses \citep{hodgeskluck2021}, which are a factor $\sim$20 higher than those for synchrotron ($\rm B_{CMB}=3.25\cdot(1+z)^2$).

\subsection{X-ray diffuse emission }
\label{sec:xray}

As shown in Fig. \ref{fig:overview} and discussed in \cite{nanni2018} and \cite{gilli2019}, the \textit{Chandra} image of the system reveals several components of emission, in both the soft (0.5-2 keV) and the hard (2-7 keV) bands. In particular, component A shows approximately an equal amount of soft and hard X-ray emission, and was suggested to be mostly ascribed to gas shock-heated by the passage of the Eastern jet, possibly combined with a contribution from CMB IC scattering. Component B, mostly visible in the hard band, is consistent with a non-thermal origin related to jet/core of the FRII radio galaxy. Components C, instead, is detected in the soft band and consistent with a thermal origin. Finally, beyond the Western hot-spot, there is a tentative detection ($\rm \sim2\sigma$) of a further spot of soft x-ray emission marked as D, which is also consistent with a thermal origin. In the light of the new radio data, here we discuss these results further.

\subsubsection{Component A}
\label{sec:acomp}

\cite{gilli2019} investigated the origin of the X-ray emission observed in component A. The X-ray spectrum was fitted with both a purely thermal model and a power-law model, which returned a gas temperature equal to >5 keV and a photon index equal to $\Gamma=1.64_{-0.35}^{0.39}$, respectively. Unfortunately due to the low number counts, the fit statistics was not found to be conclusive for neither of the two fits. 
However, when comparing the observed X-ray emission with the available radio data available, a thermal origin interpretation was preferred over the non-thermal one.

The new LOFAR and JVLA images show now much more clearly that the radio emission of the Eastern lobe entirely overlaps with component A and that component B overlaps with the core and the well-visible Eastern jet (see Fig. \ref{fig:chandra}). The confirmation of this spatial coincidence reinforces the argument for a physical connection between the radio and X-ray emission. Moreover, the spectral index in the hot-spot of the Western lobe of the radio galaxy is equal to $\rm \alpha_{inj}=0.6\pm0.1$, suggesting that the particle energy injection index in the radio galaxy is p$\sim$2.2 (p=2$\rm\alpha+1$), which is consistent with the power-law photon index $\rm \Gamma=1.64_{-0.35}^{+0.39}$ (with $\rm \Gamma=\alpha+1$) derived from the X-ray spectral analysis in the IC scenario \citep{gilli2019}.

Using our new estimates for the source volume and spectral index, as well as new considerations on the magnetic field based on results from the literature, we revise the calculations presented by \citealp{gilli2019} as shown below.

We estimate the magnetic field $B_{me}$ of the Eastern lobe using the minimum-energy assumption  following Eq. 2 in \cite{miley1980} and applying the revision proposed by \cite{brunetti1997}, to account for the contribution from lower energy electrons, which might be missed by integrating over frequency. The equations are reported below: 

\begin{equation}
\begin{aligned}
B_{\mathrm{me}}=5.69 \times 10^{-5}\left[\frac{(1+ k)}{\eta}(1+z)^{3+\alpha}\right.& \frac{1}{\theta_{x} \theta_{y} \,l \sin ^{3 / 2} \phi} \\
\left.\times \frac{S_{obs}}{\nu_{obs}^{-\alpha}} \frac{\nu_{2}^{ 0.5-\alpha}-\nu_{1}^{ 0.5-\alpha}}{0.5-\alpha}\right]^{2 / 7} \text { Gauss, }
\end{aligned}
\end{equation}

\begin{equation}
{B_{me,rev}=D(p)\gamma_{min}^{\frac{2(2-p)}{p+5}}B_{me}^{\frac{7}{p+5}}}
\end{equation}

\noindent where the symbols have the following meaning:
\\
\\
$\theta_x , \theta_y$ = angular dimensions of the radio source in arcsec along the x and y axis
\\
$k$ = ratio of energy in the heavy particles to that in electrons
\\
$\eta$ = filling factor of the emitting region
\\
$l$ = pathlength through the source in kpc
\\
$S_{obs}$ = radio flux density (in Jy) of the region at the observed frequency $\nu_{obs}$ (in GHz)
\\
$\alpha$ = spectral index
\\
$\nu_1 , \nu_2$ = lower and upper cut off frequencies presumed for the radio spectrum (in GHz)
\\
$\phi$ = angle between magnetic field and line of sight
\\
$p=2\rm\alpha+1$ = the particle energy index distribution
\\
$\gamma_{min}$ = minimum Lorentz factor of the particle distribution
\\\\

We assume $k=1$, $\eta=1$, $\phi=90^o$, $\nu_1=0.01$ GHz, $\nu_2$=100 GHz following \cite{gilli2019}. For the flux density, we use the value reported in Table~\ref{tab:flux} equal to $\rm S_{obs}=$ 35 mJy at $\nu_{obs}$=144 MHz. For the size we use $\theta_x$=22 arcsec and $\theta_y$=34 arcsec, as measured from the LOFAR image using the 3$\sigma$ contours as a reference, and we assume that the radial direction $l$ is equal to the transverse direction $\theta_x=l=$190 kpc. Based on the LOFAR-JVLA spectral index map now available (see Fig. \ref{fig:spec}), we set the spectral index to $\alpha$=0.6 (and thus $p=2.2$), which is the flattest value observed in the Western hot-spot and reasonably corresponds to the injection index value of the plasma within the lobes. $D(\gamma)=1.01$ and $\gamma_{min}=20$ are set following \cite{brunetti1997}. 
According to this, the revised equipartition magnetic field value is $B_{me,rev}=4 \ \mu G$.

We note, however, that in recent years an increasing number of studies in literature (e.g. \citealp{ croston2005, kataoka2005, migliori2007, isobe2011, ineson2017, turner2018}) have shown that magnetic fields in FRII radio galaxies are typically a factor of 2 - 3 below the values derived based on minimum energy arguments. If this were also the case for J103025+052430, we would then expect that the actual magnetic field value is in the range $B_{final}=$1.3-2 $\mu G$.

Assuming these last values, we can then compute the X-ray emission expected from IC scattering of CMB photons by the relativistic electrons (IC-CMB) in the Eastern lobe using Eq. 11 of \citet{harris1979}:

\begin{equation}
\label{eq:xflux}
S_X=\frac{(5.05 \times 10^4)^{\alpha} C(\alpha) G(\alpha)(1+ z)^{3+\alpha} S_r\nu_r^\alpha}{10^{47}B^{1+\alpha}\nu_X^\alpha},
\end{equation} 

Here $S_X$ and $S_r$ (see Table \ref{eq:fluxerr}) are the X-ray and radio flux densities in cgs units and $B$ is the magnetic field in Gauss. C($\alpha$) and G($\alpha$) are slowly varying functions of $\alpha$ (see, e.g., \citealt{harris1979} and \citealt{pacholczyk1970}): we assumed G($\alpha$)=0.521 
(it goes from 0.521 for $\alpha$=0.6, to 0.5 for $\alpha$=1.0) and C($\alpha$) = $1.15\times10^{31}$, which is good within $17\%$ for $0.5<\alpha<2.0$.

By posing $B=B_{final}$ and assuming that the X-ray spectrum is a power law with spectral index $\alpha=0.6$, i.e. equal to the radio injection index and also consistent within the uncertainties with what we measured with \textit{Chandra} for component A, we derived an X-ray flux in the 0.5 - 7 keV band of $f_{0.5-7keV}\sim (1.6-2.9)\times 10^{-15}$ \cgs.

This corresponds to a fraction of $\sim 45\%-80\%$ of the 0.5-7 keV flux measured from the \textit{Chandra} image for component A ($f_{0.5-7kev}\simeq3.7 \times 10^{-15}$ \cgs, \citealp{gilli2019}), suggesting that a significant fraction of the observed X-ray emission is likely originated by IC scattering with the CMB. This is further supported by the clear spatial coincidence between the X-ray component A and the Eastern lobe, as seen now by our new LOFAR observations. 

We note that the synchrotron-emitting particles that upscatter CMB photons ($\rm\nu_{CMB}$) to higher frequencies ($\rm\nu_{X}$) are expected to have Lorentz factors $\gamma$ equal to (see e.g. \citealp{blundell2006}):

\begin{equation}
\begin{aligned}
\gamma=\left(\left(\frac{\nu_X}{\nu_{CMB}}+\frac{1}{3}\right)\cdot\frac{3}{4}\right)^{0.5}
\end{aligned}
\end{equation}

By assuming that it is the photons from the peak of the CMB distribution that are IC upscattered by the radio-emitting electrons in the lobes/jets, then the Lorentz factors of the particles responsible for the emission observed at 1 keV is $\rm\gamma\sim$800. For magnetic fields in the range $B_{final}=$1.3-2 $\mu G$ this corresponds to particles emitting at $\sim$a few tens of MHz, below the observed radio range. The use of a spectral index equal to $\alpha=0.6$, closer to injection, for Eq. \ref{eq:xflux} is then the best approximation we can make for the energy distribution of the upscattered particles.

\subsubsection{Component C}
\label{sec:ccomp}

As already mentioned in Sect. \ref{sec:spec}, the X-ray component C overlaps with the backflow of the Western radio lobe and might represent a signature of the lobe/ICM interaction. This spot of emission extends for about 20 arcsec and is detected in the full and soft band (with a significance of $S/N\sim 2.4$ and 3.4, respectively) but not in the hard band. Despite the overall low-detection significance, we are confident that component C is real, as it is also visible in the XMM-Newton observation of the field presented by \cite{nanni2018}. The X-ray spectrum of component C is well fitted by a thermal model with T$\simeq$1 keV, while a power-law fits returns implausibly high photon indices equal to $\Gamma\sim$ 4-5, supporting a thermal origin \citep{gilli2019}.

Here we further investigate the pressure balance of this component with respect to its surrounding environment. By fixing the metallicity of the hot emitting plasma to $0.3\times$ solar (e.g. \citealp{balestra2007}), its temperature was estimated to be $kT=0.63^{+0.28}_{-0.17}$ keV \citep{gilli2019}. Assuming that the plasma is distributed within a sphere of 85 kpc radius (i.e. 10 arcsec projected radius as measured from the \textit{Chandra} image), with the measured source temperature and normalization of the best fit plasma model ({\sc APEC} within XSPEC), we derived a particle density within component C of $n_e=2.3(\pm 0.7) \times10^{-2}$cm$^{-3}$.
For reference, this is about $6\times$ higher than what was derived for component A by \cite{gilli2019}, assuming that all its X-ray emission was of thermal origin. The measured density in the hot gas bubble C translates into a thermal pressure of $P\sim n_e k T= 2.3\times 10^{-11}$ erg cm$^{-3}$. 

We note that, even assuming a mild metallicity evolution with redshift, the assumed value of $0.3\times$ solar is reasonable for the redshift of the source equal to z=1.7 (see e.g. \citealp{vogelsberger2018, flores2021}). Our result on the bubble pressure balance remains unchanged also if a more significant decrease of metals with redshift is considered. Indeed by assuming a more conservative value of $0.1\times$ solar, we computed that T would decrease by only $\sim$7\% and the electron density would increase by $\sim$30\%, which are well within the measurement errors.

It is then interesting to compare whether this bubble is in pressure equilibrium with the ambient medium around it. In \cite{gilli2019}, they assumed that most of the protocluster medium is in the form of atomic gas with temperature $T<10^5 \; K$ and column density $N_H\sim \rm 10^{19-20} cm^{-2}$, since a similar medium is observed in other $z\sim2$ protoclusters (e.g.  \citealp{cucciati2014, hennawi2015}). It is easy to show that the pressure in bubble C would largely exceed the pressure of such a medium. 

However, the cool medium pressure estimated in \cite{gilli2019} is likely a lower limit to the total pressure of the ambient medium, which could be dominated by that of a tenuous, warm-hot medium with $T=10^{5-7}$~K that is notoriously difficult to reveal \citep{nicastro2018}. 
Under the conservative hypothesis that such medium pervades the protocluster and contains most baryons, we derived its particle density as $n_{ICM}= M_{ICM}/(m_pV)$, where V is the protocluster volume, $m_p$ is the proton mass, and $M_{ICM}$ is the enclosed ICM mass. The latter can be estimated as $M_{ICM}=0.17 \times M_{DM}$, where $M_{DM}$ is the total dark matter mass in the protocluster and 0.17 is the universal baryon fraction \citep{spergel2007}.

\cite{damato2020} measured the dark matter mass enclosed within a volume of 3.9 Mpc$^3$ around the FRII radio-galaxy. This volume corresponds to the sky area covered by the ALMA observations presented in that paper ($\sim 2$ sq. arcmin) times the radial separation given by the maximum redshift interval between member galaxies ($\Delta z=0.012$). By means of both the velocity dispersion of member galaxies and the measured overdensity level, the authors estimated an enclosed dark matter mass of $M_{DM}\sim 6\times10^{13}M_{\odot}$. 

With the above numbers in hand, we derived a particle density for the ambient medium of $n_{AM}=1.1\times 10^{-4}$ cm$^{-3}$, and, in turn, a ratio between the thermal pressure of bubble C and that of the surrounding ambient medium of $P_{bubC}/P_{AM}=1.6\times 10^9\;T_{ICM}(K)$. This implies that, even assuming an extremely hot ambient medium with temperature of $10^7$ K, the bubble of gas C is still highly overpressurised and therefore it is rapidly expanding. In this process it will likely release its energy, contributing to the overall heating of the surrounding intracluster medium. 

The timescale on which bubble C is expected to cool owing to bremsshtrahlung emission is \citep{mo2010, costa2014, gilli2017}:

\begin{equation}
\rm t_{cool} \approx 8\times10^{5} \cdot (T/10^{10})^{0.5} \cdot (10^{-3}/n_e) \ Myr
\end{equation}

where T and $n_e$ are the gas temperature in K and the gas density in $\rm cm^{-3}$, respectively. For $n_e$=0.023 $\rm cm^{-3}$ and kT = 0.63 keV (T = 7.3$\rm\times10^{10}$ K) as measured in bubble C, $t_{cool} \sim$ 0.9 Gyr.
The bubble cooling time is then much longer than the estimated AGN lifetime ($\sim$70 Myr, \citealp{gilli2019}), and the bubble has then likely expanded adiabatically thus far. However, $t_{cool}$ is shorter than the $\sim$10 Gyr it will take for the structure to evolve into a massive, >1$\times\rm10^{14} \ M_{sun}$ cluster by z=0 \citep{damato2020}. By then, the bubble will have radiated away all its energy ($\rm E_{th}=3n_eVkT \sim 5\times10^{60}$ erg) , part of which may couple with the ambient medium and contribute to the overall ICM heating together with PdV work and mixing (e.g. \citealp{bourne2019}).

\section{Discussion and conclusions}
\label{sec:concl}

With its broad multi-band coverage, the giant radio galaxy J103025+052430 at z=1.7, represents an exceptional laboratory where to test how AGN jet feedback influences the growth of galaxies at high redshift and the evolution of the ICM before virialisation. In this paper we have reported new LOFAR observations at 144 MHz of the system, which we have used in combination with already published VLA observations at 1.4~GHz \citep{damato2021, damato2022} and 0.5-7 keV Chandra observations \citep{gilli2019}.

The large angular (and thus physical) size of the source gave us the possibility to perform a resolved radio spectral index analysis in the frequency range 144-1400 MHz (see Sect. \ref{sec:spec}), a very unique opportunity for a source at such high redshift. The oldest observable, radio-emitting particles have ages $\sim$8 Myr, suggesting that the jets have been active for at least this amount of time.

The observed spectral index distribution is overall consistent with expectations for FRII radio galaxies, but shows also an unexpected pattern: there is a hint for spectral flattening at the edge of the backflow in the Western lobe. 
In the backflow, particles are expected to lose their energy with time and to show steeper and steeper spectral index values moving away from the site of acceleration (the hot-spot). The detection of this flattening can thus only be explained with a scenario where particles at the edge of the backflow are interacting with the external medium and get re-energised. 

Interestingly, the end of the backflow happens to be co-spatial (at least in projection) with the X-ray component C, which was shown to have a thermal origin by \citealp{gilli2019}. What is the source of heat for component C is unsure. One natural hypothesis would be that the X-ray emitting gas is heated and compressed by the passage of the jet and in particular by the expanding radio lobe backflow. It is important to stress that because it is overpressurised with respect to the ambient medium (see Sect. \ref{sec:ccomp}), bubble C will likely expand and deposit its energy in the surrounding gas on a timescale of $\sim$0.9 Gyr, thus contributing to the overall ICM heating, well before virialisation.

While much less significant than component C, component D, located just beyond the position of the Western hot-spot (see Fig. \ref{fig:overview}) and also consistent with a thermal origin, could also represent a region where the ambient medium gets compressed by the expanding jet. If this is the case one may wonder why, given its position close to the hot-spot (where the jet pressure on the ICM is strongest), its X-ray luminosity is lower than that of component C. One simple possibility is that, since the X-ray luminosity goes as the density squared, a density difference by only a factor of 1.4 between the two components could justify the observed luminosity difference. Such a density difference is not implausible considering that the system is likely still not virialised and its baryon content is likely inhomogeneous.

Concerning the Eastern lobe, as already mentioned in Sect. \ref{sec:acomp}, \cite{gilli2019} proposed that the X-ray emission in component A was the result of gas shock-heated by the jet/lobe. Based on our new LOFAR and JVLA observations, we revised the estimate of the CMB IC scattering contribution to the X-ray emission cospatial with the Eastern lobe and found that it can actually account for most ($\sim$ 45\%-80\%) of the total 0.5-7 keV measured flux. This seem to cut back the role of the shock scenario previously proposed but, according to our numbers, it is not excluded that a non-negligible fraction of the total observed emission is still of thermal origin related to the ICM. 

Observations of CMB-IC emission associated with (giant) radio galaxies at z>1 is not unusual (e.g. \citealp{overzier2005, erlund2006, johnson2007, erlund2008, laskar2010, tamhane2015, hodgeskluck2021, carilli2022, tozzi2022}). In a few cases claims have also been made for the presence of IC ghosts, i.e. IC emission associated to old lobes of radio galaxies, having become invisible at radio frequencies due to the quick energy losses \citep{fabian2003, fabian2009, mocz2011}. Detecting the thermal emission from the ICM is instead much more challenging, especially on scales of hundreds of kpc. In the case of J103025+052430 disentangling its contribution from the IC contribution to the total observed X-ray emission would require high-resolution, high sensitivity observations, as for example in the case of the Spiderweb field \citep{carilli2022, tozzi2022}.

We note that, even if the origin of the X-ray emission in component A is entirely non-thermal, the conclusions about feedback drawn by \cite{gilli2019} do still hold, as star formation in the galaxies m1-m4 (see Fig. \ref{fig:overview}) may be promoted by the non-thermal pressure of the expanding lobe and the shocks associated to them. 

Overall, the protocluster J103025+052430 with its central radio galaxy is one of the best characterised systems and one of nicest examples of jet AGN feedback on scales of hundreds of kpc at z>1.5. Based on the estimated overdensity volume, the mass of the system is $\geq 3\times10^{13} \msun$ \citep{gilli2019, damato2020} so it will likely evolve into a galaxy cluster with M$\rm_{sys}$>$\rm10^{14} \msun$ at the z=0 \citep{damato2020}. It thus provides us with a unique chance to look back in the history of nearby rich clusters, such as the famous Perseus and Abell 2256, and surely deserves further investigation.

Future observations in both the radio and X-ray band are now extremely needed to confirm the current scenario. On the one hand, deeper X-ray observations would allow us to detect the diffuse emission at higher S/N, investigating the morphology of the brightest patches (e.g., components A, C, including any possible sharp drop in brightness), confirming the presence of the faintest ones (component D), and possibly discover new, even fainter patches. Moreover, they would enable a more robust spectral analysis that will distinguish between the thermal or non-thermal nature of component A, and place tighter constraints to the temperature of components C and D. Currently, the only instrument that may provide such an improvement is \textit{Chandra}, as it features the sharp, $\sim$arcsec angular resolution and the sensitivity required to detect faint diffuse X-rays after removal of point source contaminants. Nonetheless, this
requires significant time investments ($\gtrsim$Msec), as the effective area of \textit{Chandra} is relatively small ($\sim600$ cm$^2$ at 1 keV in the first Cycles). In the future, next generation X-ray missions such as the Survey and Time-domain Astrophysical Research eXplorer (\textit{STAR-X}\footnote{\url{http://star-x.xraydeep.org/}}), a Medium Explorer mission selected by NASA for Phase A study, and the Advanced X-ray Imaging Satellite (\textit{AXIS}, \citealt{mushotzky2019,marchesi2020}), a probe-class mission proposed to NASA, will have the sufficient 1-2 arcsec angular resolution, and the large effective area ($\sim3-10\times$ that of \textit{Chandra}) required to investigate this and other high-z protoclusters with moderately deep exposures (tens to hundreds of ksec), allowing population studies, and hence opening a new discovery space for structure formation.

From the radio point of view, new observations at complementary frequencies and a factor of a few higher in resolution, are needed to produce more detailed spectral index maps and confirm the compression scenario proposed based on the observed spectral index trend in the Western lobe. LOFAR observations at 55$\sim$144 MHz with the international stations reaching a resolution $\lesssim$1 arcsec and observations with the upgraded Giant Meterwave Radio Telescope (GMRT, \citealp{gupta2017}) at 400-700 MHz with a resolution of a few arcsec are planned to pursue this goal. Faraday rotation and (de-)polarisation analysis at frequencies $\gtrsim$1.4 GHz could also provide interesting information on the magneto-ionic properties of the ICM surrounding the radio galaxy and their interplay (e.g. \citealp{carilli2002b, guidetti2011, anderson2022}) and ALMA/ACA observations could be used to investigate the presence of any SZ signal and thus to more robustly constrain the pressure of component C (e.g. \citealp{abdulla2019, dimascolo2021}).

\begin{acknowledgements} 

M. Brienza would like to thank Giulia Migliori for the useful discussions. MB acknowledges financial support from the Italian L'Oreal UNESCO "For Women in Science" program, from the ERC-Stg ``DRANOEL", no. 714245 and from the ERC-Stg ``MAGCOW", no. 714196. MB and IP acknowledge support from INAF under the SKA/CTA PRIN “FORECaST”. IP acknowledges support from INAF under the PRIN MAIN STREAM “SAuROS” projects. We acknowledge support from the agreement ASI-INAF n. 2017-14-H.O and from the PRIN MIUR 2017PH3WAT “Blackout”. KI acknowledges support by the Spanish MCIN under grant PID2019-105510GB-C33/AEI/10.13039/501100011033. FV acknowledges the computing centre of INAF - Osservatorio Astrofisico di Catania, under the coordination of the WG-DATI of LOFAR-IT project, for the availability of computing resources and support.

LOFAR, the Low Frequency Array designed and constructed by ASTRON, has facilities in several countries, which are owned by various parties (each with their own funding sources), and are collectively operated by the International LOFAR Telescope (ILT) foundation under a joint scientific policy. The ILT resources have benefited from the following recent major funding sources: CNRS-INSU, Observatoire de Paris and Universit\'e d'Orl\'eans, France; BMBF, MIWF-NRW, MPG, Germany; Science Foundation Ireland (SFI), Department of Business, Enterprise and Innovation (DBEI), Ireland; NWO, The Netherlands; the Science and Technology Facilities Council, UK; Ministry of Science and Higher Education, Poland; The Istituto Nazionale di Astrofisica (INAF), Italy.

Part of this work was carried out on the Dutch national e-infrastructure with the support of the SURF Cooperative through grant e-infra 160022 \& 160152. The LOFAR software and dedicated reduction packages on \url{https://github.com/apmechev/GRID_LRT} were deployed on the e-infrastructure by the LOFAR e-infragrop, consisting of J.\ B.\ R.\ Oonk (ASTRON \& Leiden Observatory), A.\ P.\ Mechev (Leiden Observatory) and T. Shimwell (ASTRON) with support from N.\ Danezi
(SURFsara) and C.\ Schrijvers (SURFsara). The J\"ulich LOFAR Long Term Archive and the German LOFAR network are both coordinated and operated by the J\"ulich Supercomputing Centre (JSC), and computing resources on the supercomputer JUWELS at JSC were provided by the Gauss Centre for supercomputing e.V. (grant CHTB00) through the John von Neumann Institute for Computing (NIC).

This research made use of the University of Hertfordshire
high-performance computing facility and the LOFAR-UK computing facility located at the University of Hertfordshire and supported by STFC (ST/P000096/1), and of the Italian LOFAR IT computing infrastructure supported and operated by INAF, and by the Physics Department of Turin University (under an agreement with Consorzio Interuniversitario per la Fisica Spaziale) at the C3S Supercomputing Centre, Italy.

This research made use of APLpy, an open-source plotting package for Python hosted at \url{http://aplpy.github.com}. 

\end{acknowledgements}

\bibliographystyle{aa}
\bibliography{j1030.bib}

\end{document}